\documentclass[aps,prd,
preprintnumbers,groupedaddress]{revtex4}
\usepackage{amsmath}
\usepackage{bm}

\begin{document}
\preprint{\vtop{
{\hbox{YITP-14-94}\vskip-0pt
}
}
}


\title{$\bm{X(3872)\rightarrow J/\psi\pi\pi\pi}$   
as a Three-Step Decay} 
\author{
Kunihiko Terasaki   
}
\affiliation{
Yukawa Institute for Theoretical Physics, Kyoto University,
Kyoto 606-8502, Japan 
}

\begin{abstract}
{
Rate for the $X(3872)\rightarrow J/\psi\pi\pi\pi$ decay is studied by 
assuming that it proceeds as 
$X(3872)\rightarrow J/\psi\omega\rightarrow J/\psi\pi\rho\rightarrow 
J/\psi\pi\pi\pi$. 
The result is compared with the $X(3872)\rightarrow J/\psi\pi\pi$ decay. 
} 
\end{abstract}

\maketitle

$X(3872)$ was discovered in the $J/\psi\pi\pi$ channel~\cite{X-discovery}, 
and then its $J/\psi\pi\pi\pi$~\cite{omega-pole-exp} decay was observed. 
($J/\psi$ is written as $\psi$ hereafter.)
In accordance with experiments~\cite{X-discovery,rho-pole}, the $\pi\pi$ 
state in the $X(3872)\rightarrow \psi\pi\pi$ decay arises from the $\rho$ 
meson with the isospin $|\bm{I}| = 1$. 
However, no signal of charged partners of $X(3872)$ has been 
observed~\cite{Babar-X-charged-partner,psi-pipi-Choi}, so that it is 
considered as an iso-singlet state. 
This implies that the $X(3872)\rightarrow \psi\pi\pi$ decay is isospin 
non-conserving. 
Here, it should be noted that strength of isospin non-conserving 
hadronic interactions is of the second order of the electromagnetic 
ones~\cite{Dalitz}.   
Therefore, we expect that the ratio of branching fractions 
$R_{3\pi/2\pi} = {Br(X(3872)\rightarrow\psi\pi\pi\pi)}
                     /{Br(X(3872)\rightarrow\psi\pi\pi)}$ 
is much larger than unity, i.e., very crudely $R_{3\pi/2\pi} \sim O(\alpha^{-2})$,  
where $\alpha$ is the fine structure constant. 
However, its measured values are compiled as~\cite{compiled} 
\begin{equation}
R_{3\pi/2\pi}^{\rm exp} = 0.8\pm 0.3.                 \label{eq:exp-ratio-3pi/2pi}
\end{equation}
The above ratio implies that a large enhancement of the isospin 
non-conserving $X(3872)\rightarrow\psi\pi\pi$ decay is needed, unless the 
isospin conserving $X(3872)\rightarrow\psi\pi\pi\pi$ decay is strongly 
suppressed. 

In this short note, therefore, we study rates for the 
$X(3872)\rightarrow\psi\pi\pi$ and $X(3872)\rightarrow\psi\pi\pi\pi$ decays.  
Before going to these decays, however, we visit 
$\omega\rightarrow\pi^0\gamma$ and $\omega\rightarrow\pi\pi\pi$ decays 
for later convenience. 
Under the vector meson dominance (VMD)~\cite{VMD}, it is considered that 
the $\omega\rightarrow\pi^0\gamma$ decay proceeds as 
$\omega\rightarrow\pi^0\rho^0\rightarrow\pi^0\gamma$, and its rate is 
given by 
\begin{equation}
\Gamma(\omega\rightarrow\pi^0\gamma) 
= \frac{|g_{\omega\rho^0\pi^0}|^2}{12\pi}
\biggl|\frac{X_\rho(0)}{m_\rho^2}\biggr|^2|\bm{k}|^3,    
                                                           \label{eq:rate-for-omg-pi-gamma}
\end{equation}
where $g_{\omega\rho^0\pi^0}$ is the $\omega\rho^0\pi^0$ coupling 
constant, $X_\rho(0)$ is the $\gamma\rho^0$ coupling strength on the 
photon mass-shell and $\bm{k}$ is the photon momemtum in the rest frame 
of $\omega$. 
By assuming a $\rho$ pole dominance in the $\omega\rightarrow\pi\pi\pi$ 
decay, i.e., $\omega\rightarrow\pi\rho\rightarrow\pi\pi\pi$, and the isospin 
symmetry in strong interactions, 
it can be treated as a two-step one~\cite{Williams} and its rate is written as   
\begin{equation}
\Gamma(\omega\rightarrow\pi\pi\pi) 
= 9\int_{(s_\rho)_{min}}^{(s_\rho)_{max}} \frac{ds_\rho}{\pi} 
\Bigl\{\Gamma(\omega\rightarrow\pi^0\rho^0)\Bigl[
\frac{m_\rho\Gamma(\rho^0\rightarrow\pi^+\pi^-)}
                                                       {(m_\rho^2 - s_\rho)^2}\Bigr]\Bigr\}, 
                                                                    \label{eq:rate-for-omg-3pi}
\end{equation}
where $\Gamma(\omega\rightarrow\pi^0\rho^0)$ and 
$\Gamma(\rho^0\rightarrow\pi^+\pi^-)$ are the rates for the 
$\omega\rightarrow\pi^0\rho^0$ and 
$\rho^0\rightarrow\pi^+\pi^-$ decays, respectively, in which $\rho^0$ is off 
its mass-shell and are provided by 
\begin{eqnarray}
&&
\Gamma(\omega\rightarrow\pi^0\rho^0) 
=  \frac{|g_{\omega\rho^0\pi^0}|^2}{12\pi}|\bm{p}_{\pi^0}|^3,    
                                                          \label{eq:rate-for-omg-rho-pi}\\
&&\hspace{-3mm}
\Gamma(\rho^0\rightarrow\pi^+\pi^-) 
= \frac{|g_{\rho^0\pi^+\pi^-}|^2}{6\pi m_\rho\sqrt{s_\rho}}|\bm{p}_{\pi^+} |^3.   
                                                                  \label{eq:rate-for-rho-pi-pi}
\end{eqnarray}
Here, the momentum of $\pi^0$ in the rest frame of $\omega$ is gven by 
$|\bm{p}_{\pi^0}| = {\sqrt{\lambda(m_\omega^2,s_\rho,m_\pi^2)}}/{(2m_\omega)}$ 
with $\lambda(x,y,z) = x^2 + y^2 + z^2 - 2xy - 2yz - 2zx$ 
and the center-of-mass (c.m.) momentum of $\pi^+$ in the 
$\rho^0\rightarrow\pi^+\pi^-$ decay is provided by
$|\bm{p}_{\pi^+}| = \sqrt{s_\rho - 4m_\pi^2}/2$. 
$s_\rho$ is the invariant mass square of $\rho^0$, and  its minimum and 
maximum values in (\ref{eq:rate-for-omg-3pi}) are given by 
$(s_\rho)_{min} = (2m_\pi)^2$ and $(s_\rho)_{max} = (m_\omega - m_\pi)^2$, 
respectively. 

When $\rho$ is on its mass-shell, $s_\rho = m_\rho^2$ in 
(\ref{eq:rate-for-rho-pi-pi}) and 
$\Gamma(\rho\rightarrow\pi\pi) = 149.1$ MeV from \cite{PDG14}. 
Thus, we obtain $|g_{\rho^0\pi^+\pi^-}| = 5.975$. 
To estimate $|g_{\omega\rho^0\pi^0}|$, we use (\ref{eq:rate-for-omg-3pi}) in 
which a finite-width correction to the $\rho$ meson propagator is needed 
because it is broad. 
We here take a kinematically improved correction,     
\begin{eqnarray}
&&
(m_\rho^2 -s_\rho)^2\rightarrow (m_\rho^2 - s_\rho)^2 
+ [\sqrt{s_\rho}\,\Gamma_\rho(s_\rho)]^2, \quad 
\Gamma_\rho(s_\rho) = \Gamma_{0\rho}
\Bigl(\frac{|\bm{p}_{\pi^+}|}{|\bm{p}_{0\pi}|}\Bigr)^3
\Bigl(\frac{m_\rho}{\sqrt{s_\rho}}\Bigr), 
                                                        \label{eq:corr-of-rho-propagator}
\end{eqnarray}
by analogy with that to the Breit-Wigner form in \cite{PDG14}.  
Here, $\bm{p}_{0\pi}$ is the pion c.m. momentum in the 
$\rho\rightarrow\pi\pi$ decay and $\Gamma_{0\rho}$ is the full width of 
$\rho$, where $\rho$ is on its mass-shell. 
In the $\omega\rightarrow\pi^0\gamma$, this type of correction is not 
considered, because $s_\rho = 0$ and $m_\rho^2\gg \Gamma_{0\rho}^2$. 
Inserting the above value of $|g_{\rho^0\pi^+\pi^-}|$, the central values of the 
$\gamma\rho^0$ coupling strength 
$|X_{\rho}(0)| = 0.033\pm 0.003$ GeV$^2$~\cite{VMD-KT}, and the measured 
masses and decay rates~\cite{PDG14} into 
(\ref{eq:rate-for-omg-pi-gamma}) and (\ref{eq:rate-for-omg-3pi}) with 
(\ref{eq:rate-for-omg-rho-pi}), (\ref{eq:rate-for-rho-pi-pi}) and 
(\ref{eq:corr-of-rho-propagator}), 
we obtain  
$|g_{\omega\rho^0\pi^0}| = 12.6$ GeV$^{-1}$ from
$\Gamma(\omega\rightarrow\pi^0\gamma)$ and $14.8$ GeV$^{-1}$ from  
$\Gamma(\omega\rightarrow 3\pi)$,
where the central values of 
$\Gamma(\omega\rightarrow\pi^0\gamma)_{\rm exp} = (701\pm 25)$ keV and  
$\Gamma(\omega\rightarrow 3\pi)_{\rm exp} = (7.57\pm 0.09)$ MeV 
from \cite{PDG14} have been taken. 
Deviation between the above values of $|g_{\omega\rho^0\pi^0}|$ is small 
(roughly 10 per cent), and therefore their average value 
$|g_{\omega\rho^0\pi^0}| = 13.7$ GeV$^{-1}$ is taken hereafter. 
In this case, the central values of the measured rates for 
the $\omega\rightarrow\pi\pi\pi$, $\omega\rightarrow\pi^0\gamma$ and 
$\rho^{\pm,0}\rightarrow\pi^{\pm,0}\gamma$ decays 
are reproduced within about $20$ per cent deviations.  
Here, it has been considered that the $\rho\rightarrow\pi\gamma$ proceeds 
as $\rho\rightarrow\pi\omega\rightarrow\pi\gamma$ under the VMD and the 
central value of the $\gamma\omega$ coupling strength 
$|X_{\omega}(0)| = 0.011\pm 0.001$ GeV$^2$ on the photon 
mass-shell~\cite{VMD-KT} has been taken. 
The above result means that the $\rho$ pole contribution saturates 
the $\omega\rightarrow \pi\pi\pi$ decay in a good approximation.

We now study the isospin non-conserving $X(3872)\rightarrow\psi\pi\pi$ 
decay. 
We assume that the isospin non-conservation in the decay is caused by the 
$\omega\rho^0$ mixing~\cite{rho-pole,omega-rho-KT}, i.e., 
$X(3872)\rightarrow\psi\omega\rightarrow\psi\rho^0
\rightarrow\psi\pi^+\pi^-$, 
because it plays an essential role in isospin non-conserving nuclear forces, 
as is well known~\cite{nuclear-forces}. 
It is compatible with the experimental suggestion that the $\pi\pi$ state in 
the decay arises from $\rho$. 
In this way, we treat the decay as a two-step one and obtain its rate as 
\begin{equation}
\Gamma(X(3872)\rightarrow\psi\pi^+\pi^-) 
= \int_{(s_\rho)_{min}}^{(s_\rho)_{max}} \frac{ds_\rho}{\pi}\Bigl\{ 
\Gamma(X\rightarrow\psi\rho^0)
                      \Bigl[\frac{m_\rho \Gamma(\rho^0\rightarrow\pi^+\pi^-)}
                                 {(m_\rho^2 - s_\rho)^2}\Bigr]\Bigr\}, 
                                                                 \label{eq:rate-for-x-psi2pi}
\end{equation}
where $s_\rho$ is the invariant mass square of $\rho^0$ with its maximum 
and minimum values $(s_\rho)_{max} = (m_X - m_\psi)^2$ and 
$(s_\rho)_{min} = (2m_\pi)^2$, respectively, in (\ref{eq:rate-for-x-psi2pi}). 
($X$ denotes $X(3872)$ in the above and hereafter.) 
The rate $\Gamma(X\rightarrow\psi\rho^0)$ in which $\rho^0$ is off its 
mass-shell is given by 
\begin{equation}
\Gamma(X\rightarrow\psi\rho^0)  
= \frac{|A(X\rightarrow\psi\rho^0)|^2}{12\pi m_X^2} |\bm{p}_\psi|
\Bigl\{\frac{\lambda(m_X^2,m_\psi^2,s_\rho)}{4m_X^2} 
+ \frac{\lambda(m_X^2,m_\psi^2,s_\rho)}{4m_\psi^2} + 3s_\rho\Bigr\} 
                                                                 \label{eq:rate-for-x-psi-rho} 
\end{equation}
and $\Gamma(\rho^0\rightarrow \pi^+\pi^-)$ has the same form as 
(\ref{eq:rate-for-rho-pi-pi}). 
In the above equation, $|\bm{p}_\psi|$ is the momentum of $\psi$ in the rest 
frame of $X(3872)$ and $|A(X\rightarrow\psi\rho^0)|$ 
is given by 
$|A(X\rightarrow\psi\rho^0)|
= {|g_{X\psi\omega}||g_{\omega\rho^0}|}/{(m_\omega^2 - s_\rho)}$   
with the $X\psi\omega$ coupling strength $g_{X\psi\omega}$ and 
the $\omega\rho^0$ mixing parameter $g_{\omega\rho^0}$. 
The above formula of rate for the 
$X(3872)\rightarrow\psi\omega\rightarrow\psi\rho^0
\rightarrow\psi\pi^+\pi^-$  
has the standard form of rate for a two-step decay, but it is smaller by a 
factor two compared with the previous one~\cite{omega-rho-KT}. 
Therefore, we here revise the old formula as follows; 
the first denominator ``$2304\pi^3$'' in the right-hand-side of Eq.~(22) in 
\cite{omega-rho-KT} should be read as ``$4608\pi^3$''. 

Study of the $X(3872)\rightarrow\psi\pi\pi\pi$ decay is now in order. 
Because the $\pi\pi\pi$ state in the $X(3872)\rightarrow \psi\pi\pi\pi$ 
decay arises from $\omega$ in accordance with 
experiments~\cite{omega-pole-exp}, we assume that $X(3872)$ decays into 
$\psi\omega$ and the intermediate $\omega$ into $\pi\pi\pi$ through the $\rho$ meson pole as discussed before, i.e., 
$X(3872)\rightarrow\psi\omega\rightarrow\psi\pi\rho
\rightarrow\psi\pi\pi\pi$. 
Under the isospin symmetry, the decay can be treated as a three-step one  
and its rate is given by   
\begin{eqnarray}
&& \hspace{-10mm}
\Gamma(X(3872)\rightarrow\psi\pi\pi\pi) = 
9\Gamma(X(3872)\rightarrow\psi\omega\rightarrow\psi\pi^0\rho^0
\rightarrow\psi\pi^0\pi^+\pi^-)
\nonumber\\
&&
= 9\int_{(s_\omega)_{min}}^{(s_\omega)_{max}} \frac{ds_\omega}{\pi} 
\int_{(s_\rho)_{min}}^{(s_\rho)_{max}} \frac{ds_\rho}{\pi}\Bigl\{
\Gamma(X\rightarrow\psi\omega)
              \Bigl[\frac{m_\omega \Gamma(\omega\rightarrow \pi^0\rho^0)}
                                 {(m_\omega^2 - s_\omega)^2}\Bigr]
                     \Bigl[\frac{m_\rho \Gamma(\rho^0\rightarrow\pi^+\pi^-)}
                                 {(m_\rho^2 - s_\rho)^2}\Bigr]\Bigr\}, 
                                                        \label{eq:rate-x-psipipipi}
\end{eqnarray}
where $\Gamma(\rho^0\rightarrow\pi^+\pi^-)$ has the same form as 
(\ref{eq:rate-for-rho-pi-pi}), and 
\begin{eqnarray}
&& \hspace{-10mm}
\Gamma(X\rightarrow\psi\omega)             
= \frac{|g_{X\psi\omega}|^2}{12\pi m_X^2}|\bm{p}_\psi|
\Bigl\{\frac{\lambda(m_X^2,m_\psi^2,s_\omega)}{4m_\psi^2}
+ \frac{\lambda(m_X^2,m_\psi^2,s_\omega)}{4m_X^2} + 3s_\omega\Bigr\}, 
\label{eq:rate-x-psiomega}  
\\
&&\hspace{-2mm}
\Gamma(\omega\rightarrow \pi^0\rho^0) 
= \frac{|g_{\omega\rho^0\pi^0}|^2}{12\pi}\frac{\sqrt{s_\omega}}{m_\omega}
|\bm{p}_{\pi^0}|^3.
\label{eq:rate-omega-pirho}                                     
\end{eqnarray}
In the above, $\omega$ and $\rho$ are off their mass-shell. 
$s_\omega$ and $s_\rho$ are the invariant mass squares of $\omega$ and 
$\rho$, respectively, with their minimum and maximum values 
$(s_\omega)_{min} = (3m_\pi)^2$, $(s_\omega)_{max} = (m_X - m_\psi)^2$, 
$(s_\rho)_{min} = (2m_\pi)^2$, $(s_\rho)_{max} = (\sqrt{s_\omega} - m_\pi)^2$ 
in (\ref{eq:rate-x-psipipipi}). 

We now compare the above results with the measured ratio 
$R_{3\pi/2\pi}^{\rm exp}$ in (\ref{eq:exp-ratio-3pi/2pi}). 
By taking a ratio of the decay rates of $X(3872)$ calculated in the above, 
the unknown parameter $g_{X\psi\omega}$ can be canceled, so that our 
result is free from adjustable parameters. 
We again consider finite-width corrections to propagators of unstable 
particles. 
In the case of $\rho$ with a broad width, we take the same form as 
(\ref{eq:corr-of-rho-propagator}). 
For the propagator of $\omega$, we do not consider any kinematical 
improvement in its finite-width correction, i.e., 
$[m_\omega^2 - s_\omega\,({\rm or}\,\,s_\rho)]^2 \rightarrow 
[m_\omega^2 - s_\omega\,({\rm or}\,\,s_\rho)]^2 
+ (m_\omega\,\Gamma_{0\omega})^2$, 
because $\omega$ is narrow and $(m_\omega\Gamma_{0\omega})^2$ is 
considerably smaller than  
$[m_\omega^2 - (s_\omega)_{max}\,({\rm or}\,\,(s_\rho)_{max})]^2$, 
as long as the $X(3872)\rightarrow\psi\pi\pi$ and 
$X(3872)\rightarrow\psi\pi\pi\pi$ decays are considered, 
where $\Gamma_{0\omega}$ is the full width of $\omega$ on its mass-shell. 
The $\omega\rho^0$ mixing parameter $|g_{\omega\rho^0}|$ can be estimated 
by using the $\omega\rightarrow\pi^+\pi^-$ decay. 
When it is assumed that the decay proceeds through the $\omega\rho^0$ 
mixing, i.e., $\omega\rightarrow\rho^0\rightarrow\pi^+\pi^-$, its rate is given 
by 
\begin{equation}
\Gamma(\omega\rightarrow\pi^+\pi^-)
= \frac{|g_{\omega\rho^0}g_{\rho^0\pi^+\pi^-}|^2}{(m_\rho^2 - m_\omega^2)^2}
\frac{|\bm{p}_\pi|^3}{6\pi m_\omega^2}.               \label{eq:rate-for-omeg-pipi}
\end{equation}
Taking $s_\rho = m_\omega^2$ in (\ref{eq:corr-of-rho-propagator}), inserting 
the result into (\ref{eq:rate-for-omeg-pipi}) and comparing the resulting rate 
with its measured one  
$\Gamma(\omega\rightarrow\pi^+\pi^-)_{\rm exp} = (0.13\pm 0.01)$ 
MeV~\cite{PDG14}, 
we obtain $|g_{\omega\rho^0}| = (3.5\pm 0.2)\times 10^{-3}$ GeV$^2$.  
This value is a little bit larger than (but is still compatible with) our previous 
estimate in which any kinematical improvement in the finite-width correction 
was not considered~\cite{omega-rho-KT}. 
(This seems to mean that the kinematical improvements in the finite width 
corrections are not very important in this short note.) 
In this way, values of all the parameters which are included in the rario of 
rates $R_{3\pi/2\pi}$ are determined. 
Thus, we obtain $R^{\rm th}_{3\pi/2\pi}$ which is much larger than the 
measured one, 
$R^{\rm th}_{3\pi/2\pi} \simeq  30 \gg R^{\rm exp}_{3\pi/2\pi} = 0.8\pm 0.3$.  
This implies that the enhancement of the rate for the isospin non-conserving 
$X(3872)\rightarrow\psi\pi\pi$ cannot overcome a gap of the order 
$\alpha^2$ (in rate), though the rate is actually enhanced because of 
$m_\omega\simeq m_\rho$. 
Therefore, some unknown mechanism(s) to enhance more strongly the 
isospin non-conserving $X(3872)\rightarrow\psi\pi\pi$ and/or to suppress 
drastically the isospin conserving $X(3872)\rightarrow\psi\pi\pi\pi$ would be 
needed, if the measured ratio $R^{\rm exp}_{3\pi/2\pi}$ in 
(\ref{eq:exp-ratio-3pi/2pi}) is literally accepted. 

In summary, we have seen that the $\rho$ meson pole contribution saturates 
the $\omega\rightarrow \pi\pi\pi$ decay in a good approximation.  
Then, the $X(3872)\rightarrow\psi\pi\pi$ and 
$X(3872)\rightarrow\psi \pi\pi\pi$ decays have been studied, by assuming 
that they proceed as 
$X(3872)\rightarrow\psi\omega\rightarrow\psi\rho^0\rightarrow\psi\pi\pi$ 
(through the $\omega\rho^0$ mixing) and 
$X(3872)\rightarrow\psi\omega\rightarrow\psi\pi\rho
\rightarrow\psi\pi\pi\pi$ (because of the $\rho$ pole dominance in the 
$\omega\rightarrow\pi\pi\pi$)
and that strong interactions satisfy the isospin symmetry as usual. 
As the result, our previous formula of the rate for the 
$X(3872)\rightarrow\psi\pi\pi$ decay has been revised, and a formula of the 
rate for the $X(3872)\rightarrow\psi\pi\pi\pi$ decay as a three-step one 
has been provided. 
After that, we have compared our result on the ratio 
$R_{3\pi/2\pi}^{\rm th}$ with $R^{\rm exp}_{3\pi/2\pi}$, and have seen 
$R_{3\pi/2\pi}^{\rm th} \gg R^{\rm exp}_{3\pi/2\pi}$. 
This implies that some unexpected mechanism(s) to enhance more strongly 
the isospin non-conserving $X(3872)\rightarrow\psi\pi\pi$ and/or to 
suppress drastically the isospin conserving $X(3872)\rightarrow\psi\pi\pi\pi$ 
would be needed, if the measured ratio $R^{\rm exp}_{3\pi/2\pi}$ is literally 
accepted. 
In other words, it is not very easy to reproduce the measured ratio 
$R^{\rm exp}_{3\pi/2\pi}$, under the plausible assumptions that the $\pi\pi$ 
and $\pi\pi\pi$ states in the $X(3872)\rightarrow\psi\pi\pi$ and 
$X(3872)\rightarrow\psi\pi\pi\pi$ decays arise from the $\rho$ and 
$\omega$ mesons, respectively, as suggested by experiments, that 
the isospin symmetry works well in the strong interactions as usual and 
that the isospin non-conservation in the $X(3872)\rightarrow\psi\pi\pi$ is 
caused by the $\omega\rho^0$ mixing which plays an essential role in isospin 
non-conserving nuclear forces as is well known.  
Therefore, search for a solution to these problems is left as one of our future 
subjects, and further experimental studies on these decays are awaited. 

\section*{Acknowledgments}      
The author would like to thank Prof. S.~Takeda and Dr. D.~Sato, High Energy 
Physics group, Kanazawa University for their discussions and 
encouragements. 
He also would like to appreciate Prof. H.~Kunitomo, Yukawa Institute for 
Theoretical Physics, Kyoto University for carefull reading of the manuscript.  


\end{document}